\def\rfr#1{Eq. (\ref{#1})}

\def\dert#1#2{\frac{{{d}}{#1}}{{{d}}{#2}}}              

\def\bar{\begin{eqnarray}}
\def\ear{\end{eqnarray}}

\def\eqi{\begin{equation}}
\def\eqf{\end{equation}}
\def\eqia{\begin{eqnarray}}
\def\eqfa{\end{eqnarray}}
\def\rp#1#2{{#1\over#2}}
\def\ct#1{\cite{#1}}
\def\lb#1{\label{#1}}

\def\oc2{$\mathcal{O}(c^{-2})$}

\documentclass{ws-ijmpd}

\begin{document}

\title{PERSPECTIVES IN  MEASURING THE PPN
PARAMETERS $\beta$ and $\gamma$ IN THE EARTH'S GRAVITATIONAL FIELD
TO HIGH ACCURACY WITH CHAMP/GRACE MODELS}

\author{LORENZO IORIO}

\address{Viale Unit$\grave{a}$ di Italia 68, 70125 Bari (BA), Italy
\\e-mail: lorenzo.iorio@libero.it}

 \maketitle

\begin{history}
\received{Day Month Year}
\revised{Day Month Year}
\comby{Managing Editor}
\end{history}

\begin{abstract}
The current bounds on the PPN parameters $\gamma$ and $\beta$ are
of the order of $10^{-4}-10^{-5}$. Various missions aimed at
improving such limits by several orders of magnitude have more or
less  recently been proposed like LATOR, ASTROD, BepiColombo and
GAIA. They involve the use of various spacecraft, to be launched
along interplanetary trajectories, for measuring the post-Newtonian effects
induced by the solar gravity on the propagation of electromagnetic
waves. In this paper we
investigate the requirements  needed to
measure the
combination $\nu=(2+2\gamma-\beta)/3$ entering the post-Newtonian
Einstein pericenter precession $\dot\omega$ of a test particle
to an accuracy of the order of $\approx 10^{-5}$ with a pair of drag-free
spacecraft in the Earth's gravitational field. It turns out that
the latest gravity models from the dedicated CHAMP and GRACE
missions would allow to reduce the systematic error of
gravitational origin just to this demanding level of accuracy. In
regard to the non-gravitational errors, the spectral noise density
of the drag-free sensors required
 amounts to $10^{-8}-10^{-9}$ cm s$^{-2}$ Hz$^{-\rp{1}{2}}$
over very low frequencies. Although not yet obtainable with the
present technologies, such level of compensation is much less
demanding than those required for, e.g., LISA. As a by-product, an
independent measurement of the post-Newtonian gravitomagnetic
Lense-Thirring effect with a $\approx 1\%$ accuracy would be possible as
well. The forthcoming Earth gravity models from CHAMP and GRACE
will further reduce the systematic bias of gravitational origin in both of
such  tests.

\end{abstract}



\section{Introduction}
\subsection{The PPN parameters $\gamma$ and $\beta$ }
In the Parameterized Post-Newtonian (PPN) formalism\ct{Wil93},
developed by Nordtvedt and Will to make easier the comparison of
metric theories of gravity with each other and with experiment,
the Eddington-Robertson-Schiff parameters $\beta$ and $\gamma$
describe how much non-linearity is present in the superposition
law for gravity and how much spatial curvature is produced by a
unit mass, respectively; in general relativity $\beta=\gamma=1$.
For a recent overview of the measurements of such PPN parameters
see Refs.~\refcite{Wni05,Wil06}.

The most accurate and clean determinations of $\gamma$ come from
the effects involving the propagation of electromagnetic signals:
in their PPN expressions only $\gamma$ is present, at first order.
For the Shapiro time delay\ct{Sha64}, recently tested with
the Doppler tracking of the Cassini spacecraft, the most accurate
result is\ct{Ber03} $\gamma-1=(2.1\pm 2.3)\times 10^{-5}$,
although the accuracy of such measurement has recently
been somewhat questioned  in Ref.~\refcite{Kop07}; according to
its authors, the motion of the Sun around the Solar System barycenter,
if not properly accounted for, would induce a systematic bias of $\delta\gamma=1.2\times
10^{-4}$.

With $\beta$ the situation is less favorable because it enters the
post-Newtonian equations of motion of test particles in
conjunction with $\gamma$ itself in a rather complicated way. For
example, in the post-Newtonian expression of the well known
Einstein precession\cite{Ein15} of the pericentre $\omega$ of a test particle the combination $\nu=(2+2\gamma-\beta)/3$ is
present. E.V. Pitjeva in Ref.~\refcite{Pit05} recently processed more than 317 000 Solar
System observations (1913-2003) for the construction of the
Ephemerides of Planets and the Moon EPM2004 determining, among
other things, $\beta$ and $\gamma$ with an accuracy of $10^{-4}$.
The same limit for $\beta$ can also be obtained by
combining the Cassini result for $\gamma$ with the Lunar Laser
Ranging test\ct{Mul06} of the  Nordtvedt\ct{Nor68a,Nor68b}
effect\footnote{It is a violation of the strong equivalence
principle which is zero in general relativity. }, expressed in terms of $\eta=4\beta-\gamma-3$: indeed, from\ct{Mul06}
$\delta\eta=7\times 10^{-4}$, it
follows $\delta\beta=1.8\times 10^{-4}$. Note that if
$\delta\gamma=1.2\times 10^{-4}$ is assumed\ct{Kop07}, the error in $\beta$  becomes $\delta\beta=2.0\times
10^{-4}$.

Various theoretical models involving scalar-tensor scenarios\ct{Dam93a,Dam93b,Dam96a,Dam96b,Dam02a,Dam02b}
 predicts deviations
from unity for $\beta$ and $\gamma$ at a $10^{-6}-10^{-7}$ level,
so that it is very important to push the accuracy of such PPN
tests towards this demanding accuracy.

The proposed LATOR mission\ct{Tur06}  has, among its
goals, the determination of $\gamma$ with an accuracy of
$10^{-9}$ from the first order and a direct and independent
measurement of $\beta$ at a $\sim 0.01\%$ level via the
second-order gravity-induced deflection of  light. The expected
accuracy in the proposed radio science experiments to be conducted
with the future Mercury orbiter BepiColombo
(http://sci.esa.int/science-e/www/area/index.cfm?fareaid=30) is of
the order of\ct{Mil02} $10^{-6}$ for $\beta$ and $\gamma$.
The accurate measurement of $\beta$ and $\gamma$ is one
of the scientific goals also of the proposed mission ASTROD\ct{Ni02,Ni04}; the expected accuracy is\ct{Ni04} $1\times
10^{-9}$ for both\footnote{See also Table 5 of Ref.~\refcite{Wni05} and Ref.~\refcite{Wni08}.} . The astrometric mission
GAIA should be able to reach\ct{Vec03} $\delta\gamma\sim 10^{-7}$  through the deflection of light.
\subsection{Aim of the paper}
A satisfactorily empirical corroboration of a pillar of physics like
general relativity requires that as many independent experiments as possible are
conducted by different scientists in different laboratories. Now,
all the performed or proposed high-precision tests of post-Newtonian gravity (in the weak-field and slow-motion approximation) are
based on the effects induced by the gravitational field of the Sun
only. It is as if many independent experiments aimed to measure
fundamental physical effects were conducted always in the same
laboratory. Thus, it is worthwhile to try to use different
laboratories, i.e. other gravitational fields, to perform such
tests, even if their outcomes should be less accurate than those
conducted in the Sun's field.

In this paper we will investigate if it is possible to
realistically reach the $10^{-5}-10^{-6}$ level of accuracy in
measuring $\nu$ with a couple of dedicated drag-free spacecraft in
the terrestrial gravitational field. This analysis is motivated by the fact that, in regard to the
systematic errors of gravitational origin, reaching a $10^{-5}$ level would
 be right now possible thanks to the recent improvements in our
knowledge of the classical part of the Earth's gravitational field due to
the dedicated CHAMP
(http://www.gfz-potsdam.de/pb1/op/champ/index$\_$CHAMP.html) and
GRACE (http://www.gfz-potsdam.de/pb1/op/grace/index$\_$GRACE.html
and http://www.csr.utexas.edu/grace/) missions. Much more demanding would be to reduce the non-gravitational errors down to the required level. As we
will see, the drag-free technologies needed to implement such an
ambitious goal are not yet available because they should
work over observational time spans of several years; however, they would be less
demanding than those required for complex missions like LISA\ct{Bor03}.
\section{The Einstein pericenter precession}
The Einstein pericenter precession\ct{Ein15} of a test particle in the gravitational field of a static central body is
\eqi\dot\omega_{\rm GE}=\rp{3nGM}{c^2 a(1-e^2) },\lb{ein}\eqf
where $G$ is the Newtonian constant of gravitation, $M$ is the
mass of the central body which acts as source of the gravitational
field, $c$ is the speed of light  in vacuum, $a$ and $e$ are the
semimajor axis and the eccentricity, respectively, of the
test particle's orbit and $n=\sqrt{GM/a^3}$ is its Keplerian mean
motion.

As it is well known, \rfr{ein} allowed to explain the Mercury's
anomalous perihelion advance of 42.98 arcseconds per century
observed since the 19th century. Modern radar-ranging measurements
of the planetary perihelia in the solar field yielded a $10^{-3}$
accuracy\ct{Sha72,Sha76,Sha90}.

Hiscock and Lindblom in Ref.~\refcite{His79} analyzed with some details the
possibility of  measuring the Einstein pericenter precessions of
the natural satellites  of Jupiter and Saturn  which are much
larger than those of the planets. Iorio in Ref.~\refcite{Ior07} showed that, at
present, this possibility is not yet viable, despite of the first
data sets from the Cassini mission.

The possibility of measuring the Einstein perigee rate in the
Earth's gravity field with passive artificial satellites tracked
via laser ranging was preliminarily investigated by Rubincam
in Ref.~\refcite{Rub77} and Cugusi and Proverbio in Ref.~\refcite{Cug78}. Ciufolini and Matzner
in Ref.~\refcite{Ciu92} analyzed the data of the LAGEOS satellite using the GEM-L\ct{Ler85a,Ler85b} and GEM T1\ct{Mar88} Earth gravity models claiming a total accuracy of 20$\%$,
but such estimate is unrealistic both because of the too optimistic
evaluation of the impact of the classical part of the Earth
gravity field and of the treatment of the non-gravitational forces\ct{Mil87}
acting on the perigee of the LAGEOS-like satellites. Moreover, the very small eccentricity of the LAGEOS
orbit ($e=0.0045$) makes its perigee badly defined and highly
affected by the non-conservative perturbations. A more accurate
and reliable test was proposed in Ref.~\refcite{Ioretal02a} by
suitably combining the nodes of LAGEOS and LAGEOS II and the
perigee of LAGEOS II, and using the EGM96 Earth's gravity model\ct{Lem98}.
The estimated total accuracy is of the
order of 10$^{-3}$. An analysis of the use of the perigee of
LAGEOS II can be found in Ref.~\refcite{Luc03a}; the total error is
estimated to be $2\%$. Also the proposed LAGEOS-like satellite
LARES\cite{Ciu98} could be used for measuring the Einstein perigee advance\ct{Ioretal02b}.
Hiscock and Lindblom in Ref.~\refcite{His79} envisaged
the possibility of using geocentric drag-free spacecraft to
measure the Einstein perigee precession. A step further concerning
the use of dedicated terrestrial drag-free spacecraft for accurate
tests of post-Newtonian gravity can be found in Ref.~\refcite{Dam94}.
In it the authors suggest that it would be
possible to determine $\gamma$ and $\beta$ independently of each
other with an accuracy of $10^{-4}$ by measuring  $\nu$ and
$\eta$.

\section{Outline of the various sources of systematic errors}
When a satellite-based test of a prediction of general relativity
is proposed or performed, it is of the utmost importance to assess
the total error budget in a very reliable and accurate way by
accounting for all possible sources of systematic errors. Here we
intend the aliasing action of other competing classical features
of motion which may corrupt the genuine determination of the
investigated relativistic effect. In regard to this topic, the
choice of a suitable observable is very important in the sense
that it must reach a good compromise between the errors of
gravitational and non-gravitational origin whose simultaneous
reduction is often in conflict.
\subsection{The gravitational perturbations}
The gravitational errors come from the mismodelling of the static
and time-varying  components of the multipolar expansion of the
Newtonian gravitational potential of the Earth in spherical
harmonics\ct{Kau00}. Indeed, the static part of the even
($\ell=2,4,6,...$) zonal ($m=0$) harmonics $J_2, J_4, J_6,...$
induces secular precessions\ct{Kau00} on a satellite's node $\Omega$ and
perigee $\omega$
\begin{equation}
\left\{
\begin{array}{lll}
\dot\Omega_{\rm (geopot)}=\sum_{\ell\geq
2}\dot\Omega_{.\ell}J_{\ell},\\\\
\dot\omega_{\rm (geopot)}=\sum_{\ell\geq
2}\dot\omega_{.\ell}J_{\ell},
\end{array}
\right.
\end{equation}
where $\dot\Omega_{.\ell}$ and $\dot\omega_{.\ell}$ are the partial derivatives of the classical precessions with respect to the even zonal of degree
$\ell$ (see, e.g., Ref.~\refcite{Ior03a}), while the odd zonal harmonics $J_3, J_5,...$ affect the perigee
with long-period harmonic perturbations having frequencies
multiple of that of the perigee\ct{Kau00}. The secular
variations of the even  zonal  harmonics $\dot J_2, \dot J_4, \dot
J_6,... $  induce quadratic signatures on the node and the perigee\ct{Luc05,Ior06}.
There are also the long-period
harmonic perturbations of the node and perigee due to the tides\cite{Ior01a,Ior01,Pav02}
 whose
periods are determined by the lunisolar and satellite frequencies.
The most insidious aliasing signals are the even zonal secular and
quadratic advances and those harmonic signals that resemble
superimposed linear trends over observational time spans of some
years, like the $\ell=2, m=0$ constituent of the 18.6-year tide\cite{Ior01a}:
their mismodeled part can corrupt to a
significative extent the recovery of the genuine relativistic
linear trends of interest\cite{Ior01,Pav02}.

A way to overcome the problem of the biasing classical even zonal
secular trends consists in linearly combining the residuals
$\delta\Psi$ of the Keplerian orbital elements, denoted
generically as $\Psi$, affected by the classical and relativistic
linear rates \eqi \delta\dot\Psi^{(i)}=\sum_{\ell\geq
2}\dot\Psi_{.\ell}^{(i)}\delta J_{\ell}+\dot\Psi_{\rm
rel}^{(i)}\xi_{\rm rel},\ i=N,\eqf where $N$ is the number of
perigees and/or nodes employed, in order to solve for the
parameter  $\xi_{\rm rel}$, which accounts for the relativistic
effect of interest ($\xi_{\rm rel}=\nu$ in our case), and cancel
out the impact of as many even zonal harmonics as possible among
those recognized to be the most biasing ones\cite{Ciu96,Ior02a,Ior04}.
The coefficients $\dot\Psi_{.\ell}^{(i)}$
are the partial derivatives of the classical precessions of $\Psi^{(i)}$ with respect to the even zonal of degree
$\ell$. The so-obtained linear combinations are
sensitive to general relativity, irrespectively of the level of
mismodelling $\delta J_{\ell}$ of the static and time-varying
parts of the selected even zonal harmonics. It turns out that the
accuracy reached in the latest Earth gravity models from  CHAMP
and GRACE would allow right now to reduce the systematic error due
to the geopotential to $10^{-5}$ of the Einstein precession if a
suitable orbital geometry is chosen for a dedicated
spacecraft-based mission. We discuss this topic in detail in
Section \ref{gper}.
\subsection{The non-gravitational perturbations}
On the other hand, such really important benefits cannot be fully
capitalized if the non-gravitational forces\footnote{Let us recall
that their magnitude is proportional to the area-to-mass ratio
$\sigma/m$ of the satellite and depends on the interaction of the
electromagnetic radiation with the satellite's physical structure,
on its rotational state and its orbital geometry often in a very
complex and intricate way.} acting on the spacecraft like the
direct solar radiation pressure, Earth's albedo, Earth's IR
radiation, atmospheric drag,  thermal effects, etc.\cite{Mil87}
are not suitably accounted for and reduced to the same
level of the gravitational bias.  For example, in the case of the
Lense-Thirring\cite{Len18} tests with the existing
LAGEOS and the proposed LARES/OPTIS satellites\cite{Ciu98,Ioretal04,Lam04}, the obtainable accuracy is
set up to $\sim 1\%$ by the non-conservative forces acting on the
passive LAGEOS and LAGEOS II satellites which are not endowed with
any active mechanism of compensation of such kind of perturbations
\cite{Luc01,Luc02,Luc03b,Luc04,Lucetal04}; the
systematic error due to the geopotential would be much smaller
according to the latest CHAMP/GRACE-based models\cite{Ior05}.
Thus, it is apparent that only drag-free spacecraft could push the
bias due to the non-conservative forces to the same level of that
due to the gravitational perturbations. This topic will be the
subject of Section \ref{ngper}.
\section{How to reduce the gravitational perturbations}\lb{gper}
 The idea of using a pair of drag-free
Earth's artificial satellites was considered also in Refs.~\refcite{Ior03b,Ior03c,Ior03,Ior05};
in that case the
investigated feature of motion was the Lense-Thirring force which
affects the node and the perigee of a satellite (see Section
\ref{ltsec}). In Refs.~\refcite{Ior03b,Ior03c,Ior03} it
was shown that the difference of the perigees of two satellites in
supplementary orbital configuration\footnote{It means $a_1=a_2$,
$e_1=e_2$, $i_1=i_2-180$ deg. See Refs.~\refcite{Ciu86,Rie89}.}  is able to add the Lense-Thirring precessions and to cancel
out the classical even zonal precessions because of their
functional dependence on the inclination\cite{Ior03a}. The use of
active mechanisms of compensation of the non-gravitational
perturbations was suggested because of the strong impact that they
have on the perigees of LAGEOS-like satellites.

In the case of the Einstein precessions the situation is different
since they are independent of the inclination. This implies that
the difference of the perigees in the supplementary orbital
configuration is not a good observable because it would exactly
cancel the relativistic rates as well, independently of the values
of $\gamma$ and $\beta$. The linear combination approach involving
the nodes and the perigees of both the spacecraft must, thus, be
followed. Having the residuals of four measurable Keplerian
orbital elements at our disposal it is possible to design a
suitable observable
\eqi\delta\dot\omega^{(1)}+c_1\delta\dot\omega^{(2)}+c_2\delta\dot\Omega^{(1)}+c_3\delta\dot\Omega^{(2)}
=\nu \left[\dot\omega^{(1)}_{\rm GE}+c_1\dot\omega^{(2)}_{\rm
GE}\right]\lb{combi}\eqf which is affected by the Einstein
precessions and is independent of the static and time-dependent
parts of the first two even zonal harmonics of geopotential
$J_2,J_4$, and of the Lense-Thirring precessions. It turns out
that, also in this case, the supplementary orbital configuration
is not suitable because the coefficients weighing the perigees
would be equal and opposite cancelling the relativistic signature.
If different orbital parameters are chosen for the two spacecraft
the situation becomes favorable. Indeed, with the orbital
parameters of Table \ref{param}
\begin{table}[ph]
\tbl{Orbital parameters of the proposed
drag-free spacecraft (1) and (2): $a$ is in km and $i$ in deg.
Also their relativistic precessions, in mas yr$^{-1}$, are listed.
}
{\begin{tabular}{@{}ccccccc @{}} \toprule
Satellite & $a$  & $i$  & $e$ & $\dot\omega_{\rm GE}$ &
$\dot\omega_{\rm LT}$ &
 $\dot\Omega_{\rm LT}$ \\ \colrule
(1) & 13000 & 50 & 0.2 & 2955.7 & -53.0 & 27.5 \\
(2) & 12000 & 103 & 0.2 & 3610.5 & 23.6 & 35.0 \\ \botrule
\end{tabular}\label{param}}

\end{table}
%
%
%
we have for our
$J_2-J_4-$Lense-Thirring-free combination
\begin{equation}
\left\{
\begin{array}{lll}
c_1=2.2817,\\\\
c_2=-0.6861,\\\\
c_3=0.5165,\\\\
\dot\omega^{(1)}_{\rm GE}+c_1\dot\omega^{(2)}_{\rm GE}= 11194.1\
{\rm mas\  yr}^{-1},
\end{array}
\right.
\end{equation}
and the relative systematic error due to the uncanceled even
zonal harmonics amounts to
\eqi\left.\rp{\delta\nu}{\nu}\right|_{\rm geopot}\sim 3\times
10^{-5},\lb{err}\eqf if the currently available CHAMP/GRACE-based
EIGEN-CG03C Earth gravity model\cite{For05} is
adopted for the calculation. Note that \rfr{err} represents a
realistic upper bound of the gravitational bias because it has
been obtained by linearly summing the absolute values of the
individual errors up to degree $\ell=20$. Note that this level of
truncation is quite adequate because of the high altitudes of the
satellites: indeed, it turns out that the numerical value of
$\left.\delta\nu/\nu\right|_{\rm geopot}$ does not change if one
adds more even zonals. The current models from CHAMP and GRACE are
still preliminary and do not yet exploit the full data sets
gathered by such missions which are still ongoing. Thus, it is
likely that when new, more robust and reliable Earth gravity
solutions will be available the estimate of \rfr{err} will further
improve, all the satellites' orbital parameters being equal.

In regard to the time-varying gravitational errors, the periods of
the nodes and the perigees of the satellites are listed in Table
\ref{periodi}.
\begin{table}\tbl{Periods, in yr, of the node  and the
perigee of the proposed drag-free spacecraft (1) and (2)
}
{\begin{tabular}{@{}ccc@{}} \toprule

Satellite & $P(\Omega)$  & $P(\omega)$  \\
\colrule
(1) & -1.71 & 2.06\\
(2) & 3.7 & -2.22\\
\botrule

\end{tabular}\label{periodi}}

\end{table}
%
%
%
%
%
%
%
%
It can be noted that they amount to just a few years. This means
that the sinusoidal perturbations induced on the nodes and the
perigees by the $\ell=2, m=1$ constituent of the $K_1$ solar tide\cite{Ior01a}
and the odd zonal harmonic $J_3$, which have just
the same periods of the satellite's node and perigee,
respectively, and are not canceled out by the combination of
\rfr{combi}, do not represent serious bias over reasonably long
observational time spans. A potentially less favorable situation
occurs for the $\ell=3, m=1, p=1, q=-1$ oceanic constituent of the
$K_1$ tide\cite{Ior01a} which is not canceled by \rfr{combi} and
affects the perigees with perturbations whose periods are -10.0 yr
and -5.6 yr, respectively. For the complete details of the action
of the $\ell=3$, $m=1$ oceanic constituents of $K_1$ on the
perigees of (1) and (2), see Table \ref{marea}.
\begin{table}[ph]
\tbl{
Periods $P$, in yr, and mismodeled
amplitudes $\delta A$, in mas,  of the tidal perturbations induced
on the perigees of the proposed drag-free spacecraft (1) and (2)
by the $\ell=3$, $m=1$ oceanic constituents of the $K_1$ solar
tide according to the results of Table 6.4.8.5-1 of
Ref.~36.
}
{\begin{tabular}{@{}ccccc@{}} \toprule
Satellite & $P(p=1\ q=-1)$  & $\delta A(p=1\ q=-1)$ & $P(p=2\
q=1)$ &
$\delta A(p=2\ q=1)$\\
\colrule
(1) &  -10.0 & -10.1 & -0.9 & 2.1\\
(2) & -5.6 & 7.0 & 1.4 & 3.3\\
\botrule
\end{tabular}\label{marea}}
\end{table} %
According to the global ocean tide model of the EGM96 solution\cite{Lem98},
it turns out that the most insidious
tidal effect is the $K_1\ \ell=3,\ m=1,\ p=1,\ q=-1$ perturbation
on the perigee of satellite (1) with a mismodeled amplitude of
-10.1 mas. Thus, the maximum relative error, over 10 yr, on the
combination of \rfr{combi} amounts to $9\times 10^{-5}$. However,
it must be noted that such an estimate is largely pessimistic both
because it assumes a peak value after 10 yr and because it is
based on the relatively old EGM96 model: more recent and accurate
models like GOT99, CSR 4.0, FES2004 would greatly reduce the
impact of such a source of bias.

If in constructing the combination of \rfr{combi} one chose to
cancel $J_6$ instead of the Lense-Thirring precessions, and a
nearly-polar orbital configuration was adopted for both the
satellites, a bound $\left.\delta\nu/\nu\right|_{\rm
geopot}=10^{-7}$ could be reached, according to EIGEN-CG03C. But,
in this case, it would not be possible to separate\footnote{Also
the post-Newtonian $J_2/c^2$ effects due to the Earth's oblateness\cite{Sof88,Hei90}
 affect all
the combination proposed here. They must be modeled, otherwise
they would induce a systematic bias of $10^{-4}$ on the
measurement of the Einstein precession. } the Einstein precessions
from the Lense-Thirring ones which are weighted by a different
combination of $\gamma$ and $\beta$, namely $\mu=(1+\gamma)/2$.
Moreover, inclinations close to 90 deg would be harmful\cite{Ior02b}
because of the extremely long node periods and, thus, of
the $K_1$ tidal perturbations.

%
%
%
%

%
\section{The impact of the inclination errors}
Another source of systematic error is related to the uncertainty
in the knowledge of the spacecraft inclinations and their impact
on the bias due to the uncanceled even zonal harmonics: indeed,
recall that the coefficients $\dot\omega_{.\ell}$ and
$\dot\Omega_{.\ell}$ of the classical even zonal precessions\ct{Ior03a} do depend on the inclinations as well.

To this aim, we considered $\left.\delta\nu/\nu\right|_{\rm
geopot}$ as a function of the two variables $i^{(1)}$ and
$i^{(2)}$ and studied its behavior in small ranges $\Delta
i^{(1)}$ and $\Delta i^{(2)}$  around the nominal values of Table
\ref{param} which simulate the errors in the inclinations. It
turns out that for a very conservative and pessimistic assumption
$\Delta i=2$ deg for both $i^{(1)}$ and $i^{(2)}$ the relative
error in $\nu$ varies by just $9\times 10^{-6}$.


\section{The accuracy required for the non-gravitational perturbations}\lb{ngper}

In this Section we will investigate the level to which the
non-gravitational accelerations should at most affect the spacecraft  if a
relative systematic error of $10^{-6}$  was to be obtained.

The Gauss equations for the variations of the  node and the
perigee are\cite{Mil87} \bar
\dert{\Omega}{t} & = & \rp{1}{na\sin i\sqrt{1-e^2}}\ A_n\left(\rp{r}{a}\right)\sin (\omega+f),\lb{nod}\\
\dert{\omega}{t} & = & -\cos
i\dert{\Omega}{t}+\rp{\sqrt{1-e^2}}{nae}\left[-A_r\cos
f+A_t\left(1+\rp{r}{p}\right)\sin f\right],\lb{perigeo} \ear in
which  $f$ is the true anomaly counted from the pericentre,
$p=a(1-e^2)$ is the semilactus rectum of the Keplerian ellipse,
$i$ is the inclination of the orbit to the Earth's equator, $A_r,\
A_t,\ A_n$ are the radial, transverse (in-plane components) and
the normal (out-of-plane component) projections of the perturbing
acceleration $\boldsymbol{A}$, respectively, on the co-moving
frame
$\{\boldsymbol{\hat{r}},\boldsymbol{\hat{t}},\boldsymbol{\hat{n}}\}$.

We are interested in the effects averaged over one orbital period
$P$: they can be obtained by evaluating the right-hand-sides of
\rfr{nod} and \rfr{perigeo} on the unperturbed Keplerian ellipse
\eqi r=\frac{p}{1+e\cos f},\lb{kep}\eqf and averaging them with
\eqi \frac{dt}{P}=\frac{(1-e^2)^{3/2}}{2\pi(1+e\cos f
)^2}df.\lb{media}\eqf

Let us start with the node and consider the action of an
out-of-plane acceleration consisting of a part $N^{(0)}$ constant
over one orbital revolution and a part varying with the orbital
frequency \eqi A_n=N^{(0)}+N_S\sin f+N_C\cos f,\lb{nac}\eqf where
$N_S$ and $N_C$ are to be considered as constant over one orbital
revolution. Now \rfr{nac} must be inserted into \rfr{nod} and
treated as previously outlined. By using
\begin{equation}
\left\{
\begin{array}{lll}
\int_0^{2\pi}\rp{\sin(\omega+f)}{(1+e\cos f)^3}df=-\rp{3\pi
e}{(1-e^2)^{5/2}}\sin\omega,\\\\
\int_0^{2\pi}\rp{\sin(\omega+f)(N_S\sin f+N_C\cos f)}{(1+e\cos
f)^3}df=\rp{\pi}{(1-e^2)^{5/2}}\left[(1-e^2)N_S\cos\omega+(1+2e^2)N_C\sin\omega\right],\nonumber
\end{array}
\right.
\end{equation}
 one finally gets
\begin{equation}
\begin{array}{lll}
\left\langle\dert\Omega t\right\rangle = \rp{1}{2\sin
i}\sqrt{\rp{a}{GM(1-e^2)}}\left\{[(1+2e^2)N_C-3eN^{(0)}]\sin\omega+(1-e^2)N_S\cos\omega\right\}\lb{nodong}.
\end{array}
\end{equation}

The same procedure holds for the perigee. By inserting
\begin{equation}\left\{\begin{array}{lll} A_r = R^{(0)}+R_S\sin f+R_C\cos
f,\\\\
A_t = T^{(0)}+T_S\sin f+T_C\cos f\end{array}\right.\end{equation}
into the second term of the right-hand-side of \rfr{perigeo} and
using

\begin{equation}
\left\{
\begin{array}{lll}
\int_0^{2\pi}\rp{\cos f}{(1+e\cos f)^2}df=-\rp{2\pi
e}{(1-e^2)^{3/2}},\\\\
\int_0^{2\pi}\rp{(2+e\cos f)\sin f}{(1+e\cos f)^3}df =0,\\\\
\int_0^{2\pi}\rp{\cos f(R_S\sin f+R_C\cos f)}{(1+e\cos
f)^2}df=\rp{2\pi
R_C}{e^2}\left[1-\rp{1-2e^2}{(1-e^2)^{3/2}}\right],\\\\
\int_0^{2\pi}\rp{(2+e\cos f)(T_S\sin f+T_C\cos f)\sin f}{(1+e\cos
f)^3}df =-\rp{\pi
T_S}{e^2}\left[2+\rp{e^2-2}{(1-e^2)^{3/2}}\right],
\end{array}
\right.
\end{equation}
it is possible to obtain \eqi\left\langle\dert\omega
t\right\rangle=-\cos i\left\langle\dert\Omega t
\right\rangle+\left\langle\dert\omega
t\right\rangle^{(0)}+\left\langle\dert\omega
t\right\rangle^{(C/S)},\eqf with
\begin{equation}
\left\{
\begin{array}{lll}
\left\langle\dert\omega
t\right\rangle^{(0)}=\sqrt{\rp{a(1-e^2)}{GM}}R^{(0)},\lb{bizz1}\\\\
\left\langle\dert\omega
t\right\rangle^{(C/S)}=-\sqrt{\rp{a}{GM}}\rp{(1-e^2)^2}{e^3}\left\{\left[1-\rp{1-2e^2}{(1-e^2)^{3/2}}\right]R_C
+\left[1+\rp{e^2-2}{2(1-e^2)^{3/2}}\right]T_S\right\}.
\end{array}
\right.
\end{equation}
This result tells us that the perigee would be affected by secular
or long-period perturbing rates, according to the frequencies
embedded in $N^{(0)}$, $N_S$, $N_T$, $R^{(0)}$, $R_{C}$ and
$T_{S}$.

From \rfr{bizz1} it is possible to obtain the level to which the
non-gravitational perturbations should be reduced in order not to exceed a given systematic error in the measurement of the gravitoelectric
perigee precession. For $a\sim 12000-13000$ km and $e=0.2$, a
$10^{-6}$ relative accuracy could be reached with a residual
acceleration level of\footnote{For a comparison, the impact of the
direct solar radiation pressure on the existing LAGEOS satellites
amounts to\cite{Luc01} $\sim 10^{-7}$ cm s$^{-2}$.} $10^{-11}$
cm s$^{-2}$. The drag-free technologies currently under
development for, e.g., the LISA mission\cite{Bor03}
should allow a cancelation of the non-gravitational accelerations
down to $3\times 10^{-13}$ cm s$^{-2}$ Hz$^{-\rp{1}{2}}$. In our
case, with an orbital frequency of $7\times 10^{-5}$ Hz, a drag-free level of $\sim
10^{-9}$ cm s$^{-2}$ Hz$^{-\rp{1}{2}}$  would be
sufficient: such a cancelation should be reached and maintained
over time spans some years long, i.e. in a frequency range
spanning from about 0 to $10^{-5}$ Hz. It is certainly not an easy
task to reach. No long-term experience about the performance of
such systems is available at present; current high precision
accelerometers show-by extrapolating present data-a spectral noise
density of\cite{Sch05} $\sim 10^{-8}$ cm s$^{-2}$ Hz$^{-\rp{1}{2}}$. A very accurate in-orbit
calibration would be required during the entire duration of the
mission.

%
\section{The measurement errors}\lb{vacca}
For the perigee and the node the measurable quantities are
$r=ea\omega$ and $r=a\sin i\Omega$. Thus, for the combination of
\rfr{combi} the measurement error  can be written as\footnote{Please note that the labels $(1)$ and $(2)$ refer here to the two spacecraft.} \eqi\delta
r_{\rm meas } \left[
\frac{1}{a^{(1)}e^{(1)}}+c_1\frac{1}{a^{(2)}e^{(2)}}+c_2\frac{1}{a^{(1)}\sin
i ^{(1)}} +c_3\frac{1}{a^{(2)}\sin i^{(2)}}\right] \lb{vg}\eqf By
using the values of Table \ref{param} and by assuming $\delta
r_{\rm meas}=0.1$ cm over an observational time span $T=10$ yr,
one gets for the relative error
\eqi\left.\frac{\delta\nu}{\nu}\right|_{\rm meas}=2\times
10^{-6}.\eqf About the assumptions made, it must be noted that we
consider an experimental millimeter accuracy in determining the
spacecraft orbit, in a root-mean-square sense, over 10 years: it
is certainly not an easy task to be accomplished, especially in view of the fact that no empirical along-track
accelerations should be fitted: otherwise the signal of interest could be removed as well. Moreover, it must be said that attempts
to improve measurement errors are not the same as the capability of reconstructing the orbit with mm accuracy over 10 years.
%
%
\section{The measurement of the Lense-Thirring effect}\lb{ltsec}
As a by-product, our active spacecraft could also be used to make
an independent measurement of the Lense-Thirring precessions\cite{Len18,Bar74,Sof89,Ash93,Ior01b}
\begin{equation}
\left\{
\begin{array}{lll}
\dot\Omega_{\rm LT}=\mu\rp{2GS}{c^2 a^3(1-e^2)^{3/2}},\\\\
\dot\omega_{\rm LT}=-\mu\rp{6GS\cos i}{c^2 a^3(1-e^2)^{3/2}},
\end{array}
\right.
\end{equation}
where $S$ is the spin angular momentum of the central rotating
mass.
 Indeed, if one combines the nodes and the perigees so to cancel
out $J_2$, $J_4$ and the Einstein precessions, and solves for
$\mu$, the resulting combination is obtained
\eqi\delta\dot\omega^{(1)}+k_1\delta\dot\omega^{(2)}+k_2\delta\dot\Omega^{(1)}+k_3\delta\dot\Omega^{(2)}=\mu[\dot\omega_{\rm
LT}^{(1)}+k_1\dot\omega_{\rm LT}^{(2)}+k_2\dot\Omega_{\rm
LT}^{(1)}+k_3\dot\Omega_{\rm LT}^{(2)}]\eqf with
\begin{equation}
\left\{
\begin{array}{lll}
k_1=-0.9,\\\\
k_2=0.21,\\\\\
k_3=-2.7,\\\\
 \dot\omega_{\rm LT}^{(1)}+k_1\dot\omega_{\rm
LT}^{(2)}+k_2\dot\Omega_{\rm LT}^{(1)}+k_3\dot\Omega_{\rm
LT}^{(2)}=-161\ {\rm mas \ yr^{-1}}.
\end{array}
\right.
\end{equation}
In this case, the major limiting factor is represented by the
systematic error due to the uncanceled even zonal harmonics
because the drag-free level of cancelation of the
non-gravitational perturbations discussed in Section \ref{ngper}
would allow to reduce their impact well below the percent level.
It turns out that $\delta\mu/\mu|_{\rm geopot}=9\times 10^{-3}$,
according to EIGEN-CG03C. Such a result is not so good as it would
be with the supplementary orbital configuration, and it could be
reached also with only one new passive satellite and the existing
LAGEOS and LAGEOS II (see Ref.~\refcite{Ior05}). Nonetheless,
$\delta\mu/\mu|_{\rm geopot}$ should be improved by the new Earth
gravity field solutions from CHAMP and GRACE pushing the overall
accuracy of such test towards the $\sim 0.1\%$ level. This
achievement could not be reached with the low-cost alternative
involving one new satellite and the LAGEOS ones because of the
uncanceled action of the non-gravitational perturbations.
\section{Conclusions}
In this paper we investigated what is needed to measure
the combination $(2+2\gamma -\beta)/3$ of the post-Newtonian
gravitoelectric Einstein perigee precession of a test particle at
a $\approx 10^{-5}$ level of accuracy by using a couple of drag-free
spacecraft in the Earth's gravitational field.
The observable proposed
is a linear combination of the nodes $\Omega$ and
the perigees $\omega$ of such satellites to be analyzed over an
observational time span of several years. It allows to cancel out
the impact of the static and time-varying parts of the first two
even zonal harmonics of the terrestrial geopotential, which
represent a major source of systematic error, and of the
post-Newtonian gravitomagnetic Lense-Thirring precessions. By
using the latest Earth gravity model EIGEN-CG03C from CHAMP and
GRACE, the bias induced by the remaining uncanceled even zonals
amounts to $10^{-5}$, for a suitable choice of the orbital
configuration of the satellites with $a\sim 13000-12000$ km and
$i\sim 50-103$ deg. When the forthcoming, more accurate and
reliable CHAMP/GRACE-based Earth's gravity models will be
available, it should be possible to push the gravitational bias
down to $10^{-6}$. An active reduction of the impact of the
non-conservative perturbations is required in order to fully
benefit of the low systematic error of gravitational origin. It turns out that the level of
cancelation of the non-gravitational perturbations should be of
the order of $10^{-9}$ cm s$^{-2}$ Hz$^{ -\rp{1}{2} }$. Although
the currently available drag-free sensors do not yet allow to
reach such performances, also because they should be maintained
over several years, they are less demanding than those required
for, e.g., LISA. As a by-product of the proposed mission, an
independent measurement of the Lense-Thirring effect with a
$\approx 1\%$ accuracy would be possible as well. In this case, given
the drag-free level required for the previous test, the most
limiting factor is represented by the uncanceled even zonals; the
forthcoming CHAMP/GRACE models should reduce their impact on this
test as well. Another stringent and demanding requirement is reaching the ability of reconstructing
the satellites' orbits at $\approx$mm level in root-mean-square sense over a time span of 10 years.



\end{document}